\documentclass[%
 reprint,
 superscriptaddress,
 showpacs,preprintnumbers,
 amsmath,amssymb,
 aps,
 pra,
 longbibliography,
lengthcheck,%
]{revtex4-1}

\usepackage{graphicx}
\usepackage{dcolumn}
\usepackage{bm}
\usepackage{hyperref}
\usepackage{color}
\usepackage{ulem}

\begin{document}

\preprint{APS/123-QED}

\title{
Momentum-Space Electromagnetic Induction in Weyl Semimetals
}

\author{Hiroaki Ishizuka}
\affiliation{
Department of Applied Physics, The University of Tokyo, Bunkyo, Tokyo, 113-8656, JAPAN 
}

\author{Tomoya Hayata}
\affiliation{
Department of Physics, Chuo University, 1-13-27 Kasuga, Bunkyo, Tokyo, 112-8551, JAPAN 
}

\author{Masahito Ueda}
\affiliation{
Department of Physics, The University of Tokyo, Bunkyo, Tokyo, 113-0033, JAPAN 
}
\affiliation{
RIKEN Center for Emergent Matter Sciences (CEMS), Wako, Saitama, 351-0198, JAPAN
}

\author{Naoto Nagaosa}
\affiliation{
Department of Applied Physics, The University of Tokyo, Bunkyo, Tokyo, 113-8656, JAPAN 
}
\affiliation{
RIKEN Center for Emergent Matter Sciences (CEMS), Wako, Saitama, 351-0198, JAPAN
}

\date{\today}

\begin{abstract}
  We theoretically study the effect of the Berry curvature on the transport properties of Weyl semimetals in the nonadiabatic process, which results in nonlinear optical responses. In the adiabatic process, the Berry curvature, which involves the time derivative of the Bloch states, contributes to the transport properties such as the adiabatic Thouless pump. Although this effect is very weak in usual solids, it is enhanced in Weyl  semimetals, where the Berry curvature contributes to observable nonlinear optical responses due to its nodal structure.  In this paper, using semiclassical Boltzmann theory, we show that the d.c. photocurrent induced by the Berry curvature robustly persists even in the limit of short scattering time. We also show that the photocurrent is well explained as a consequence of the electromagnetic induction in momentum space. The results indicate that the electromagnetic induction gives rise to a non-dissipative photocurrent that is robust even when decoherence occurs within a time scale shorter than the periodicity of the light. We also discuss the second harmonic response of the a.c. current when the electron distribution is displaced from the ground state by an external field.
\end{abstract}

\pacs{
}

\maketitle

\section{Introduction}

The semiclassical transport theory of metals well explains many of the basic properties of them. In this theory, the properties are explained on the basis of the classical dynamics of charge carriers such as electrons or holes, in which the carrier velocity is replaced by the group velocity of the Bloch states. Recent studies on the anomalous transport phenomena, however, have revealed that the conventional transport theory is incomplete for explaining the rich transport phenomena discovered in solids. One of the key features lacking in the simplest version of the transport theory is the contribution from the Berry phase in momentum space~\cite{Karplus1954,Sundaram1999,NagaosaRMP,Xiao2010}. It was shown that the Berry phase in the orbital degrees of freedom of the Bloch electrons is related to rich physics in solids such as the quantum Hall effect~\cite{Thouless1982,Kohmoto1985} and the electric polarization in ferroelectric compounds~\cite{Resta1992,KingSmith1993}, and is now considered as one of the key features that characterize the electronic states in solids such as in topological insulators~\cite{Kane2005,Moore2007,Fu2007,Roy2009}.

On the other hand, it was also pointed out that the Berry phase of time-periodic systems induces non-dissipative particle transport in the adiabatic process~\cite{Thouless1983,Niu1984}, which was recently observed in cold atom experiments~\cite{Nakajima2016,Lohse2016}. In this phenomenon, when the system is insulating for all $t\in[0,T)$, where $T$ is the period of the time-dependent perturbation, the charge carried by the system per a cycle, $Q$, is expressed by the Berry curvature of the electron Bloch states with time derivative~\cite{Thouless1983},
\begin{eqnarray}
  Q= q\int_0^T dt \int_{BZ} \frac{dk}{2\pi} e(\bm k,t),\label{eq:thoulesspump}
\end{eqnarray}
where $q<0$ is the elementary charge. Here, the second integral in Eq.~\eqref{eq:thoulesspump} is over the first Brillouin zone, and
\begin{eqnarray}
  e(k,t)&=&-{\rm i}\left< \partial_tu(k,t)\right|\left. \partial_{k}u(k,t)\right>\nonumber\\
  &&\qquad\qquad+ {\rm i}\left< \partial_ku(k,t)\right|\left. \partial_tu(k,t)\right> \label{eq:defE1}
\end{eqnarray}
is the Berry curvature for the valence band in the $t$-$k$ space with $u(k,t)$ being the Bloch state of electrons in the valence band at wavenumber $k$ and time $t$ (Here, we assumed one-dimensional systems).

In principle, this theory is applicable to any periodic systems including solids. Indeed, in a generalized formalism of wave-packet theory, it is shown that the term identical to Eq.~\eqref{eq:defE1} appears in the equation of motion of the wave packet~\cite{Xiao2010};
\begin{eqnarray}
  \dot{r}_\mu &=& \partial_{k_\mu}\varepsilon(\bm k) - e_\mu(k,t),\\
  e_\mu(\bm k,t)&=&-{\rm i}\left< \partial_tu(k,t)\right|\left. \partial_{k_\mu}u(k,t)\right>\nonumber\\
  &&\qquad\qquad+ {\rm i}\left< \partial_{k_\mu}u(k,t)\right|\left. \partial_tu(k,t)\right>.
\end{eqnarray}
Here, $r_\mu$ ($\mu=x,y,z$) is the $\mu$th component of the position vector of the wave packet, and $\bm k=(k_x,k_y,k_z)$ is the wavenumber of the Bloch state $u(\bm k,t)$. Unfortunately, the physics related to $e_\mu(\bm k,t)$ is often considered to be negligible in bulk solid materials. The main obstacle is the energy scale of the time-dependent field~\cite{Ishizuka2016}. Since the amount of particles pumped during the adiabatic cycle is proportional to the integration of the Berry curvature, for an observable pumping, a large deformation of  the wave function is required. In solids, however, the characteristic energy scale of the materials, such as the band width, is much larger than that of the controllable time-dependent field, e.g., the electromagnetic field. Therefore, it is generally difficult to significantly deform the electronic wave function by external fields.

In a recent paper~\cite{Ishizuka2016}, however, the present authors have proposed that the effect of $e_\mu(\bm k,t)$ may give an observable contribution to the transport properties of the Weyl semimetals~\cite{Murakami2007,Wan2011,Burkov2011, Xu2011,Fang2012, Chen2013,Guan2015,Ueda2015,Tian2016,Huang2016,Chen2016, Huang2015,Weng2015,Lv2015,Xu2015,Sun2015}. In particular, using an adiabatic approximation, it was theoretically discussed that $e_\mu(\bm k,t)$ is induced by an incident circularly polarized light in doped Weyl semimetals and results in a non-dissipative photocurrent. It was further pointed out that this phenomenon could be interpreted as the electromagnetic induction in momentum space, where the circular motion of the magnetic monopoles (Weyl nodes) induces a d.c. emergent electric field.

In this paper, we study in details the transport phenomena induced  by $e_\mu(\bm k,t)$ in Weyl semimetals. Using a generalized Weyl Hamiltonian with tilting~\cite{Katayama2006,Kino2006,Kobayashi2007,Kobayashi2008} and the $k^2$ terms, we first study the distribution of $e_\mu(\bm k,t)$ around the Weyl node. We show that the circular motion of the Weyl node induces the d.c. $e_\mu(\bm k,t)$ that resembles the magnetic field induced by the circulating point charge. We then study the transport properties induced by $e_\mu$.  We show, by using the Boltzmann theory in the $\tau\omega\ll1$ limit, that the non-dissipative photocurrent studied in Ref.~\cite{Ishizuka2016} robustly persists even in the limit, where the decoherence of the quantum phase occurs much faster than the cycle of the pumping, namely, the period of $e_\mu(\bm k,t)$ induced by the incident light. From the calculation, the non-dissipative nature of the photocurrent is directly obtained; it does not depend on the relaxation time. We, however, show that in the case where the deformation of the bands by the electric field is large, the coefficient of the photocurrent changes.  This effect quantitatively changes the results, and may totally cancel out the photocurrent in a special case, as shown for the case of the tilted Weyl Hamiltonian without the $k^2$ terms. In the last, we show that  the a.c. electric current with the frequency twice as high as that of the incident light also arises in a nonequilibrium setup. Distinct from the d.c. photocurrent, this second harmonic response is a dissipative current and depends on the relaxation time.

The transport properties of Weyl fermions have recently attracted attention, from high energy physics to its condensed matter realizations named Weyl semimetals, owing to its rich physics in nonlinear electromagnetic responses, such as the chiral magnetic effect~\cite{Fukushima2008,Ebihara2016}, the nonlinear anomalous Hall effect~\cite{Sodemann2015,Chan2016_1}, the photovoltaic effects~\cite{Taguchi2016,Chan2016_2}, and the giant second harmonic generation~\cite{Wu2016}. These phenomena originate from several different mechanisms such as the chiral anomaly, spin-momentum locking, and the Berry curvature (anomalous velocity) of the Bloch wave functions. Our theory shows that another form of the Berry curvature, $e_\mu(\bm k,t)$, also contributes to the electromagnetic responses of Weyl semimetals, which are potentially relevant to the candidate materials for noncentrosymmetric Weyl semimetals, such as TaAs~\cite{Huang2015,Lv2015,Sun2015,Xu2015,Weng2015}.

The remainder of this paper is organized as follows. In Sec.~\ref{sec:model}, we introduce the generalized Weyl Hamiltonian studied in this paper. In Sec.~\ref{sec:eminduction}, we discuss the main idea of the paper, i.e, the electromagnetic induction in momentum space. We show how $e_\mu(\bm k,t)$ is induced by the circular motion of a Weyl node, and present its physical argument based on an analogy with Maxwell theory; it is similar to the generation of the magnetic field around a point charge that circulates around a point. Section~\ref{sec:optresponse} discusses the experimental signature of $e_\mu(\bm k,t)$, that is, the photocurrent and second harmonic response. Section~\ref{sec:summary} is devoted to the summary and discussions regarding other related works. In Appendix~\ref{sec:boltzmann}, we elaborate on the technical details of the Boltzmann theory used in Sec.~\ref{sec:optresponse}.

\section{Model} \label{sec:model}

\begin{figure}
  \includegraphics[width=\linewidth]{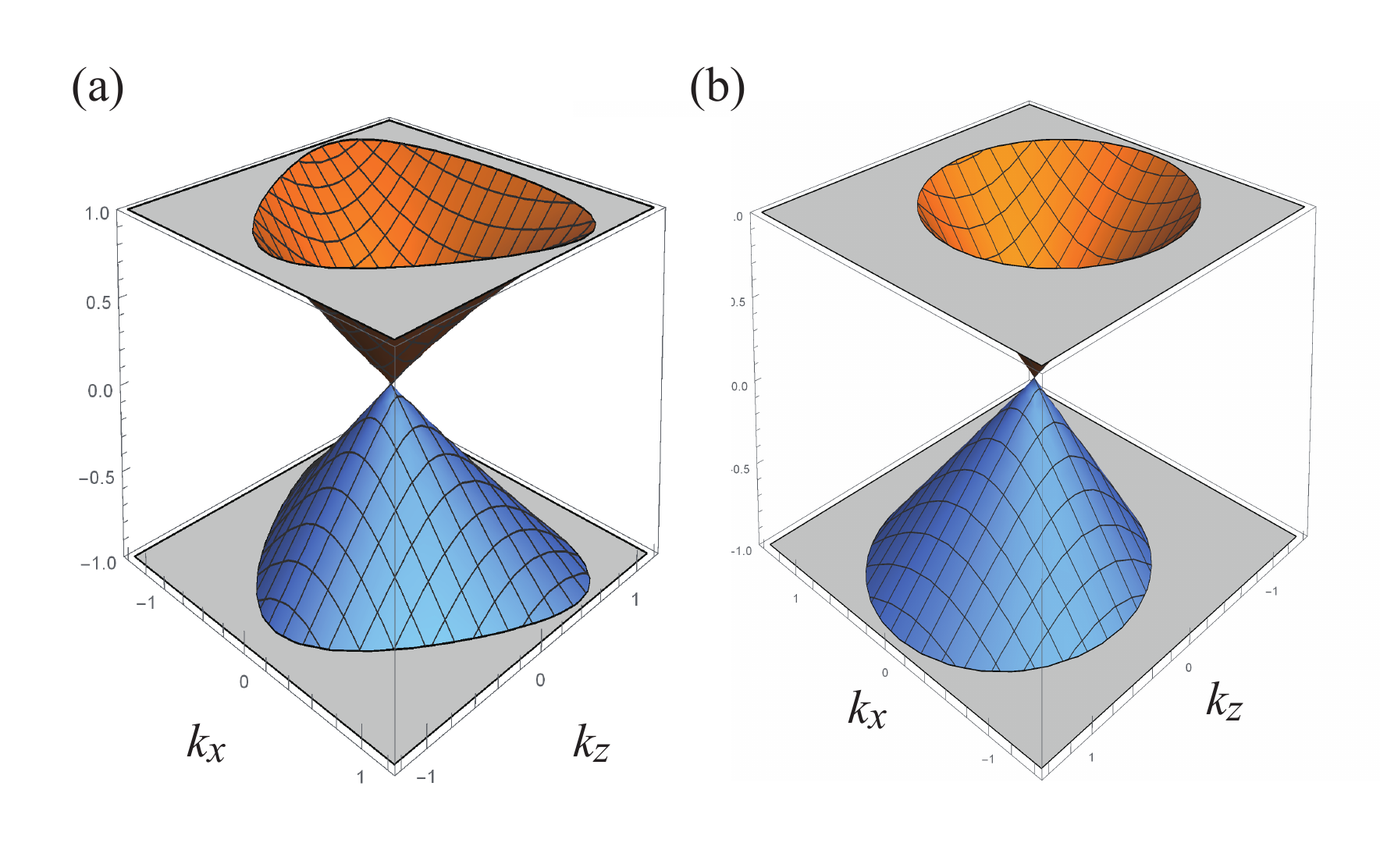}
  \caption{Dispersion of the Hamiltonian in Eq.~\eqref{eq:Rweyl} at $k_y=0$ with $v=v_z=1$ 
and $D_x=D_y=0$. (a) Weyl Hamiltonian with nonlinear term, $v_0=0$, $\alpha_1=0$, and 
$\alpha_2=-1$, and (b) with tilting $v_0=1/4$ and $\alpha_1=\alpha_2=0$.}
  \label{fig:dispersion}
\end{figure}

In this paper, we study a generalized Weyl Hamiltonian with quadratic terms and the tilting; the Hamiltonian is given by 
\begin{eqnarray}
  H(\vec{k},t)=v_0k_z + \sum_a \sigma_a R_a(\vec{k},t), \label{eq:Hweyl}
\end{eqnarray}
with
\begin{subequations}
  \begin{align}
    R_x(\vec{k})&= \pm v k_x + g D_y + \frac{\alpha_2}2 k_xk_z, \label{eq:Rweyl_a}\\
    R_y(\vec{k})&= \pm v k_y - g D_x + \frac{\alpha_2}2 k_yk_z, \label{eq:Rweyl_b}\\
    R_z(\vec{k})&= \pm v_z k_x + \frac{\alpha_1}2 (k_x^2+k_y^2-2k_z^2). \label{eq:Rweyl_c}
  \end{align}\label{eq:Rweyl}%
\end{subequations}
Here, $D_\alpha$ ($\alpha=x,y$) are the electric field along the $x$ and $y$ directions, and $k_\alpha$ ($\alpha=x,y,z$) are the displacement of the wavenumber from the nodal point; the axis of $k$ is chosen so that the pair of nodes connected by time-reversal or spatial-inversion symmetry are along the $z$ axis. The velocity of Weyl electrons is given by $v$ and $v_z$. Due to the uniaxial anisotropy about the $z$ axis, in a Weyl semimetal, the velocity along the $z$ axis, $v_z$, is generally different from that of $x$ and $y$ axes, $v$.

In this paper, as the coupling to the external electric field, we consider an electron orbital coupling allowed by symmetry [the second terms in Eqs.~\eqref{eq:Rweyl_a} and~\eqref{eq:Rweyl_b}]. The incident light propagating along the $z$ axis is given by
\begin{subequations}
  \begin{align}
    D_x&= D\cos(\omega t),\\
    D_y&= D\sin(\omega t+\chi),
  \end{align}\label{eq:Efield}%
\end{subequations}
where $D$ is the square root of the intensity of the light, $\omega$ is the frequency of the light, and $\chi$ is the phase shift; the light is circularly polarized for $\chi=0,\pi$, and linearly polarized for $\chi=\pi/2, 3\pi/2$. $g$ in Eq.~\ref{eq:Rweyl} is generally nonzero when the Weyl node is away from a symmetric point, such as the $\Gamma$ point.

In addition to these terms, we consider the tilting of the Weyl node and the quadratic terms in the dispersion relation. The tilting is given by the first term in Eq.~\eqref{eq:Hweyl}, and the quadratic terms by the third term in Eqs.~\eqref{eq:Rweyl_a} and \eqref{eq:Rweyl_b}, and the second term in Eq.~\eqref{eq:Rweyl_c}. These terms violate the symmetry of the dispersion relation under $k_z\to-k_z$. This is seen in the band dispersions shown in Fig.~\ref{fig:dispersion}. Figure~\ref{fig:dispersion}(a) shows the dispersion of the Hamiltonian with $v_0=0$, $\alpha_1=0$ and $\alpha_2=-1$, and Fig.~\ref{fig:dispersion}(b) shows that with $v_0=1/4$, $\alpha_1=0$ and $\alpha_2=0$. The nonlinear term proportional to $\alpha_2$ deforms the Fermi surface into a triangular shape, violating the symmetry under $k_z\to -k_z$. The tilting also violates the symmetry related to the transformation $k_z\to -k_z$, which deforms the Fermi surface into an oval shape. As we will discuss in Secs.~\ref{sec:eminduction} and \ref{sec:optresponse}, this deformation of the Fermi surface is necessary to realize nonzero $e_\mu(\bm k,t)$. 

\section{$\bf\it e$ field in Weyl Semimetals} \label{sec:eminduction}

\begin{figure}
  \includegraphics[width=\linewidth]{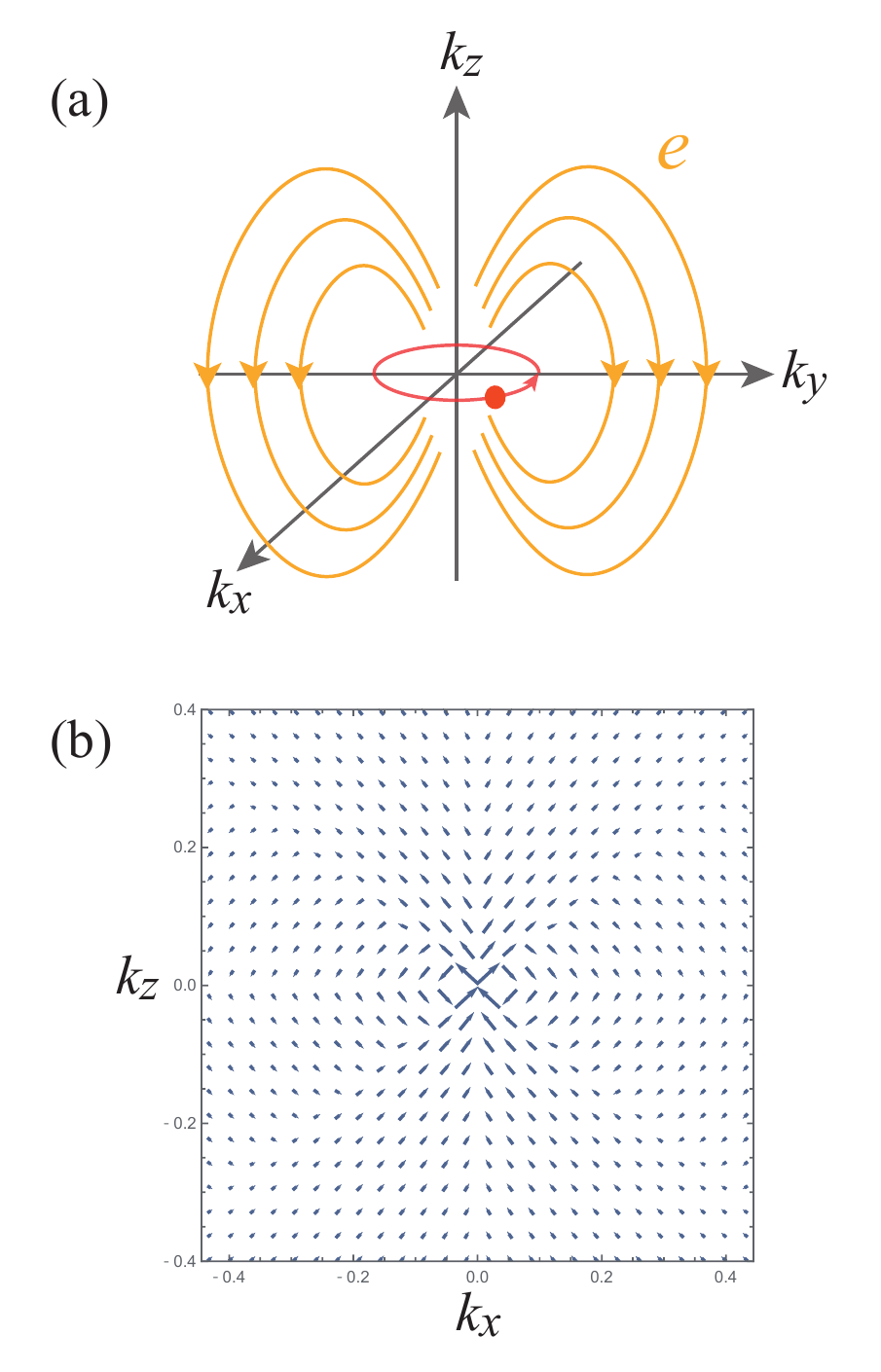}
  \caption{(a) Schematic figure of circular motion of a Weyl node (magnetic monopole) in momentum space and an induced emergent electric field. In analogy with a magnetic field induced by the circulating point charge, the emergent electric field is induced by the cyclic motion of the magnetic monopole. (b) Averaged emergent electric field per a cycle calculated using Hamiltonian in Eq.~\eqref{eq:Hweyl}. The calculation is done for $v=v_z=1$, $\alpha_{1,2}=0$, $g=0.1$, $D_x^2+D_y^2=1$, $\chi=0$, and $\omega=1$. Each arrow shows the electric field $\bm{e}$  projected onto $k_x-k_z$ plane with its length being proportional to $|\bm{e}|^{1/6}$.}
  \label{fig:weyl}
\end{figure}

In this section, we provide an intuitive argument that explains the qualitative features of the anomalous transports presented in Sec.~\ref{sec:optresponse}. We describe an analogy with the electromagnetism of a moving point charge that well explains the results of optical responses presented in the subsequent sections~\cite{Ishizuka2016}. We consider an analog to the Amp{\`e}re law described by
\begin{eqnarray}
  \nabla\times\bm H=\frac{\partial \bm E}{\partial t}+4\pi\bm j, \label{eq:maxwell}
\end{eqnarray}
where $\bm j$ is the current, and $\bm H$ and $\bm E$ are the magnetic and electric fields, respectively.

In Weyl semimetals, it is known that the Weyl nodes are the magnetic monopoles of the orbital Berry curvature $\bm b$, i.e., $\nabla_k\cdot\bm b=\rho_m\ne 0$ at the node, where $\rho_m$ is the magnetic charge density of the monopole. When the Weyl node moves dynamically in the momentum space, a similar effect to Eq.~\eqref{eq:maxwell} may appear in momentum space, which is described by
\begin{eqnarray}
  \nabla_k\times\bm e=-\frac{\partial\bm b}{\partial t}-4\pi\bm j_m, \label{eq:maxwell_mom}
\end{eqnarray}
where $\bm e$ is the emergent electric field, and $\bm j_m$ is the magnetic charge current in momentum space, satisfying the conservation law $\partial_t\rho_m+\nabla_k\cdot \bm j_m=0$. Equation~\eqref{eq:maxwell_mom} is the momentum-space analog of the electromagnetic induction, 
\begin{eqnarray}
  \nabla\times\bm E=-\frac{\partial\bm B}{\partial t}-4\pi\bm j_m,
\end{eqnarray}
and describes the dynamical generation of $\bm e$ ($\bm B$ is the magnetic flux density). For the magnetic charge, the role of electric and magnetic fields changes. Therefore the emergent electric field $\bm e$ is generated according to Eq.~\eqref{eq:maxwell_mom} by the dynamical motion of the magnetic monopole (Weyl node). The generated $\bm e$ shows the same distribution with the magnetic field $\bm H$ generated according to the Amp{\`e}re law in Eq.~\eqref{eq:maxwell} by the dynamical motion of the electric charge. In Eq.~\eqref{eq:Hweyl}, in the presence of an electric field, the position of the Weyl node shifts from its original position by $-\frac{g}vD_y$ and $\frac{g}vD_x$ in the $k_x$ and $k_y$ directions, respectively. When the light is circularly polarized, namely, for the case of $\chi=0$ in Eq.~\eqref{eq:Efield}, the Weyl node moves in a circular orbit around its original position. In analogy with the circulating point charge, the circular motion of the Weyl node induces a d.c. $e_\mu$ field that penetrates through the orbit as schematically shown in Fig.~\ref{fig:weyl}(a). 

To see whether this analogy works in momentum space, we calculate the time average of $e_\mu(\bm k,t)$:
\begin{eqnarray}
\bar{e}_\mu(\bm k) = \int_0^{\frac{2\pi}\omega}\frac\omega{2\pi}dt\; e_\mu(\bm k,t)f[\bm k,\bm\gamma(t),t].
\end{eqnarray}
Up to the order of $O(D^2)$, in the case of $\alpha_{1,2}=0$ and a circularly polarized light ($\chi=0$), the solution with $f[\bm k,\bm\gamma(t),t]=1$ reads
\begin{eqnarray}
  \bar{e}_\mu(\bm k) = \pm\omega g^2D^2v_\mu\frac{3 \sum_\nu\hat z_\nu \kappa_\nu 
\kappa_\mu - \kappa^2\hat z_\mu}{2\kappa^5},\label{eq:edist}
\end{eqnarray}
where $v_\mu=v$ for $\mu=x,y$ and $\kappa_\mu=v_\mu k_\mu$; $z_\mu$ is the $\mu$th component of the unit vector along $z$, i.e., $\hat{\bm z}=(0,0,1)$. The distribution of $\bar{e}_\mu(\bm k)$ described by Eq.~\eqref{eq:edist} is depicted in Fig.~\ref{fig:weyl}(b); it shows the $x$ and $z$ components of $\bar{e}_\mu(\bm k)$ in the $k_y=0$ plane. Due to the continuous rotational symmetry around the $z$ axis, ${\bar e}_y(\bm k)=0$ in this plane. The intensity of $\bar{e}_\mu(\bm k)$ is proportional to the frequency of rotation, $\omega$, and the square of $gD$, which is proportional to the area surrounded by the orbit of the monopole. These features have a clear analogy to the magnetic field induced by a circulating charged particle, which is proportional to the intensity of the current (velocity) and the area surrounded by the loop of the orbit.

Indeed, Eq.~\eqref{eq:edist} is identical to the magnetic field distribution for the point charge rotating in a circular motion, in the distance from the orbit of the point charge, renormalized by the velocity $v_\mu$ in Eq.~\eqref{eq:Rweyl}. For the classical electromagnetism, the magnetic field distribution for moving point charges can be exactly calculated using the Li\'enard-Wiechert potential. In the limit of large distance, it is given by replacing $gD\to l\sqrt{q/c}$ in Eq.~\eqref{eq:edist}:
\begin{eqnarray}
  \bm H_\mu = \frac{q l^2\omega}c\frac{3 \sum_\nu\hat z_\nu x_\nu x_\mu - x^2\hat z_\mu}{x^5},
\end{eqnarray}
where $x_\mu$ ($\mu=1,2,3$) is the real space coordinate, $x^2=\sum_\mu x_\mu^2$, $q$ is the charge of the point particle, $l$ is the radius of the orbit, and $c$ is the speed of light (we assumed $c\gg\omega l$ and $x\gg l$).

We remark here that the magnetic field distribution far away from the circulating point charge cancels out after the integration over the angular directions. Therefore, the total $e_\mu$ cancels out  upon the integration over the momentum space, that is,
\begin{eqnarray}
  \int \frac{d^3k}{(2\pi)^3} \bar{e}_\mu(\bm k)=0.
\end{eqnarray}
Thus, the contribution from $e_\mu$ vanishes if the Fermi level is at the node. This is consistent with the explicit calculation conducted in Sec.~\ref{sec:optresponse}. With finite doping, the integration is replaced by the volume inside the Fermi surface, and therefore the integration can be nonzero. However, when we have the symmetric Fermi surface, such as at $v_0=\alpha_1=\alpha_2=0$, the integration again cancels out. Therefore, to make the integration nonzero, it is necessary to break the symmetric structure of $e_\mu(\bm k,t)$ by deforming the Fermi surface from the symmetric one at $v_0=\alpha_1=\alpha_2=0$. We show, in Sec.~\ref{sec:optresponse}, that this idea indeed works and $e_\mu(\bm k,t)$ induces optical responses.

\section{Optical Response} \label{sec:optresponse}

To study the optical responses of the Hamiltonian in Eq.~\eqref{eq:Hweyl}, we here utilize the Boltzmann transport theory. In Sec.~\ref{sec:formalism}, we elaborate on the detail of the Boltzmann theory. The optical responses obtained from the Boltzmann theory are presented in Secs.~\ref{sec:nonlinear} and \ref{sec:tilted}. In Sec.~\ref{sec:nonlinear} we study the photocurrent and the second harmonic response induced by the emergent electric field in a Weyl Hamiltonian with nonlinear terms. Those in tilted Weyl Hamiltonians are discussed in Sec.~\ref{sec:tilted}.

\subsection{Boltzmann Theory} \label{sec:formalism}

In this paper, we focus on the electric current induced by the emergent electric field, particularly on the photocurrent and second harmonic generation. For this purpose, we employ the Boltzmann theory with the relaxation time approximation. In the Boltzmann theory, the electric current is given by 
\begin{eqnarray}
J_\mu = q\int \frac{d^3x}{(2\pi)^3} \dot{r}_\mu f[\bm k,\bm\gamma(t),t],
\end{eqnarray}
where
\begin{eqnarray}
  \dot{\bm r} = \nabla_{\bm k} \varepsilon (\bm k) - \bm b(\bm k,t)\times \dot{\bm k} - \bm e(\bm k,t)\label{eq:rdot}
\end{eqnarray}
is the time derivative of the position of the wave packet~\cite{Xiao2010}, $\varepsilon (\bm k)$ is the energy dispersion, and $\bm \gamma(t)$ is the time-dependent parameters in the Hamiltonian; in our case, it is $\bm\gamma=(gD_y,-gD_x,0)$ (we leave the zero in the third component for later convenience). The first term on the right-hand side of Eq.~\eqref{eq:rdot} is the group velocity of the wave packet, and the second term is the anomalous velocity contributed from the Berry curvature,
\begin{eqnarray}
  b_\mu(\bm k,t)=\epsilon_{\mu\nu\lambda} \partial_{k_\nu} a_\lambda(\bm k),
\end{eqnarray}
where
\begin{eqnarray}
  a_\mu(\bm k,t)=-{\rm i}\langle u(\bm k,t)|\partial_{k_\mu} u(\bm k,t)\rangle
\end{eqnarray}
is the Berry connection. This term is known to contribute to the photocurrent when the electron distribution function deviates from its equilibrium distribution~\cite{Moore2010,Sodemann2015}. In this paper, we focus on the contribution from the third term in Eq.~\eqref{eq:rdot}, which describes the Thouless pumping in the adiabatic limit:
\begin{eqnarray}
  J_\mu^{(e)} = -q\int \frac{d^3k}{(2\pi)^3} \bm e(\bm k,t)f[\bm k,\bm\gamma(t),t].
\end{eqnarray}
When the system is subjected to a monochromatic light, the induced $\bm e(\bm k,t)$ field can be decomposed in terms of the frequency of light $\omega$: 
\begin{eqnarray}
  \bm e(\bm k,t) = \sum_n\bm e(\bm k,n) e^{i n\omega t}.
\end{eqnarray}
We find both of $\bm e(\bm k,0)$ and $\bm e(\bm k,\pm2)$ are of order $O(D^2)$, while $\bm e(\bm k,\pm1)$ are of order $O(D)$. In the following, we focus on the d.c. and second harmonic responses that arise from these terms.

To evaluate the electron distribution function, $f[\bm k,\bm\gamma(t),t]$, we here use the relaxation time approximation to take into account of the modulation of the distribution function due to the time-dependent external electric field. This approximation should be valid when the change of the electron distribution is limited to a small range in the energy close to the Fermi surface, i.e., when the energy scale of the electric field is much smaller than the Fermi energy $\mu\sim vk_F$. The derivation of the distribution function is elaborated in Appendix~\ref{sec:boltzmann}. Here, we just note the main result of the appendix; the electron distribution can be approximated, by using the equilibrium distribution $f_0(\bm k)$, as
\begin{eqnarray}
f(\bm k,t)&\sim&f_0(\bm k-q\tau \bm \tilde D), \label{eq:distribution}
\end{eqnarray}
where $\bm \tilde D=(\tilde D_x,\tilde D_y,0)$ with
\begin{eqnarray}
  \tilde D_x &=& D_x-\frac{g}{vq\tau}D_y,\\
  \tilde D_y &=& D_y+\frac{g}{vq\tau}D_x,
\end{eqnarray}
when $\tau\omega\ll1$ and $\alpha_{1,2}k_F\ll v$. Here, $\tau$ is the relaxation time, and $k_F$ is the Fermi wavenumber. Therefore, the change in the Fermi surface due to the change in the band structure (the terms independent of $\tau$ in $q\tau \bm \tilde D$) appears in a manner similar to the usual electric field, but shifts its phase. The first terms on the right-hand side of the above equations come from the usual electromagnetic coupling by replacing $k$ by $k - qA$.

\subsection{Nonlinear Weyl Hamiltonian} \label{sec:nonlinear}

To study the optical responses, we first consider the case with $v_0=0$, but nonzero $\alpha_{1,2}$. This is the model considered in the previous paper, in which we studied the limit $\tau\omega\gg1$, namely, the adiabatic limit~\cite{Ishizuka2016}.  We show that the expression of the photocurrent obtained in Ref.~\cite{Ishizuka2016} is qualitatively valid even in the limit of $\tau\omega\ll1$. In addition, we show that $e_\mu$ also gives rise to the second harmonic response, i.e., the a.c. current with frequency $2\omega$, when the displacement of the electron distribution in momentum space occurs. These results indicate that, although the dynamics of the Fermi surface affects the quantitative features of the adiabatic current, most of the qualitative features robustly persist in all $\tau\omega$ regions, and are well explained in terms of the momentum-space electromagnetic induction.

We first consider the photocurrent induced by the $e_\mu$ field. We consider the short scattering time regime $\tau \omega \ll1$, so that the distribution function is approximated as in Eq.~\eqref{eq:distribution}. Since the electric current induced by the light is proportional to the time and momentum average of $e_\mu(\bm k,t)$,
\begin{eqnarray}
  \bar{e}_\mu = \int_0^{\frac{2\pi}\omega}\frac\omega{2\pi}dt\int d^3k e_\mu(\bm k,t)f_0(\bm k-q\tau \bm \tilde D),
\end{eqnarray}
we here discuss ${\bar e}_\mu$ instead of the current. To calculate the integral inside the Fermi surface, we focus on the case $\alpha_{1,2}k_F \ll v,v_z$, and expand the dispersion relation around the Fermi surface for $\alpha_{1,2}=0$ along the radial direction. The change in $\bm{k}_F$ due to $\alpha_{1,2}$, $\Delta \bm{k}$, is given by solving 
\begin{eqnarray}
  \mu-\varepsilon(\bm{k}_F^{(0)})+\Delta \bm{k} \cdot \partial_{\bm k} 
\varepsilon(\bm{k}_F^{(0)})=0,
\end{eqnarray}
where $\bm{k}_F^{(0)}$ is the Fermi surface for $\alpha_{1,2}=0$. From the fact that $\mu-\varepsilon(\bm{k}_F)\sim O(\alpha_i)$, we expect $\Delta \bm{k} \sim O(\alpha_i)$. Hence, in general, we need to consider terms up to $O(\Delta \bm{k}^2)$ to fully take into account terms up to $O(\alpha_i^2)$. However, we find that the $O(\alpha_i^2)$ contribution to $\Delta \bm{k}$ vanishes for the current model. Therefore, considering up to the linear order is sufficient for the current calculation.

We first consider the limit of $q\tau |{\bm D}|\gg\frac1v |\boldsymbol\gamma|$. 
In this limit, we find 
\begin{eqnarray}
  \bar{e}_{z\rm R,L} &=& \pm\pi\frac{4(v^2-2v_z^2)\alpha_1-3vv_z\alpha_2}
{30v^5v_z^3}\alpha_1 (\mu g D)^2\omega\cos(\chi),\label{eq:nonlinear:e1}
\nonumber\\
\end{eqnarray}
where the $+$ and $-$ signs on the right-hand side are for the right-handed ($\bar{e}_{z\rm R}$) and the left-handed ($\bar{e}_{z\rm L}$) Weyl nodes, respectively. Here, $\mu$ is the Fermi energy measured from the node. As discussed in Ref.~\cite{Ishizuka2016}, $\bar{e}_{z\rm R(L)}$ does not depend on the sign of $\mu$, i.e., the expression does not depend on whether it is electron doped or hole doped. This result is exactly the same with that in the adiabatic limit, and implies the robustness of the adiabatic current. Indeed, reflecting the adiabatic nature of the current, Eq.~\eqref{eq:nonlinear:e1} does not depend on the relaxation time.

In regard to the polarization of the light, $\chi$, we find that the photocurrent is maximized when the light is circularly polarized ($\chi=0,\pi$), while it vanishes for linearly polarized lights ($\chi=\pi/2,3\pi/2$). This polarization dependence can also be understood in terms of the electromagnetic induction; the circular motion of a magnetic monopole (Weyl node) induces the d.c. emergent electric field penetrating through the orbit, while a vibration of the monopole does not.

In the limit of $q\tau |{\bm D}| \ll\frac1v |\boldsymbol\gamma|$, the magnitude of the photocurrent is modified due to the change in the Fermi surface. We find
\begin{eqnarray}
  \bar{e}_{z\rm R,L} &=& \pm\pi\frac{16(v^2-v_z^2)\alpha_1-9vv_z\alpha_2}{30v^5v_z^3}\alpha_2 (\mu g D)^2\omega\cos(\chi).\nonumber\\
\end{eqnarray}
In this situation, however, the difference from the result in Eq.~\eqref{eq:nonlinear:e1} is only quantitative. Many of the main features in Eq.~\eqref{eq:nonlinear:e1} robustly remain, such as the polarization dependence and the non-dissipative nature (the result is independent of $\tau$).

A new feature of the optical response in the Boltzmann theory is a second harmonic response. We find that a second harmonic response is induced by $e_\mu$ when the displacement of the electron distribution occurs. In the $q\tau |{\bm D}| \gg\frac1v |\boldsymbol\gamma|$ limit it reads
\begin{eqnarray}
  \tilde{e}_{z\rm R,L}(t) &=& \pi gq\tau\frac{(6v^2-4v_z^2)\alpha_1-3vv_z\alpha_2}{15 v^4v_z^3}\alpha_2\nonumber\\
  &&\quad\times(\mu D)^2\omega\sin(\chi)\cos(2\omega t+\chi).
\end{eqnarray}
Here $\tilde{e}_z$ denotes the momentum integration of $e_z(\bm k,t)$. In contrast to the photocurrent, the second harmonic response occurs only for the linearly polarized light ($\chi=\pi/2,3\pi/2)$, and vanishes for the circularly polarized light ($\chi=0,\pi$). Another contrasting feature is that the current depends on $\tau$, which indicates that the second harmonic response is dissipative. Reflecting this feature, the second harmonic response vanishes in the $q\tau |\bm D| \ll\frac1v |\boldsymbol\gamma|$ limit.

\subsection{Tilted Weyl Hamiltonian} \label{sec:tilted}

We next consider the case with tilting ($v_0\ne0$). For simplicity, we here ignore the contribution from $\alpha_{1,2}$. In this case, the Fermi surface can be exactly calculated for $|v_0|<|v_z|$; using the polar axis $\bm \kappa=(vk\sin\theta\cos\phi,vk\sin\theta\sin\phi,v_zk\cos\theta)$, it reads
\begin{eqnarray}
  \kappa_F(\theta)=\frac{v_z\mu}{v_z+v_0\cos\theta}.\label{eq:tilted:fermi}
\end{eqnarray}

We first consider the limit of $q\tau |{\bm D}|\gg\frac1v |\boldsymbol\gamma|$. 
In this limit, $\bar{e}_{z\rm R,L}$ reads
\begin{eqnarray}
  \bar{e}_{z\rm R,L} &=& \pm\frac\pi{2v^2} X(v_0/v_z) (g D)^2\omega\cos(\chi),
\label{eq:tilted:e1gen}
\end{eqnarray}
where
\begin{eqnarray}
  X(x)=\int_0^\pi d\theta \sin\theta \left(1+3\cos2\theta\right)
\ln\left(\frac1{1+x\cos\theta}\right).\nonumber\\
\end{eqnarray}
This is, again, the same with that in the adiabatic limit. When $|v_0|\ll|v_z|$, Eq.~\eqref{eq:tilted:e1gen} becomes
\begin{eqnarray}
  \bar{e}_{z\rm R,L} &=& \pm\pi\frac{4v_0^2}{15v^2v_z^2} (g D)^2\omega\cos(\chi).
\end{eqnarray}
The result is qualitatively similar to the case with $\alpha_{1,2}\ne0$ and $v_0=0$; the photocurrent shows the maximum for a circularly polarized light, but vanishes for a linearly polarized one. Also it is proportional to $D^2$, and does not depends on $\tau$.

One distinct feature for this case appears in the doping dependence. Unlike the case with $\alpha_{1,2}\ne0$, in the presence of tilting, the intensity of the photocurrent is independent of doping. This owes to the difference in the distortion of the Fermi surface. For the case in Sec.~\ref{sec:nonlinear}, the relative amount of the distortion of the Fermi surface is proportional to $\alpha_{1,2}k_F/v\propto\mu$, while it is independent of $k_F$ in the present case, as seen in Eq.~\eqref{eq:tilted:fermi}.

We, however, note that the result in Eq.~\eqref{eq:tilted:e1gen} does not indicate a finite photocurrent without doping. In our calculation, we use the relaxation time approximation. For this approximation to hold, it is necessary to have a doping $\mu\gg vq\tau D, gD$. Therefore, the photocurrent in Eq.~\eqref{eq:tilted:e1gen} needs to be corrected when the Fermi level approaches the node. 

We also investigate the case $q\tau |{\bm D}| \ll\frac1v |\boldsymbol\gamma|$. In this case, we find that the contribution from the Fermi surface modulation cancels the adiabatic current. Therefore, no photocurrent appears.

In the last, we discuss the second harmonic response. In the limit of $q\tau |{\bm D}| \gg\frac1v |\boldsymbol\gamma|$ when $\alpha_{1,2}=0$ and $v_0\ne0$, we get
\begin{eqnarray}
  \tilde{e}_{z\rm R,L}(t) &=& -\frac\pi{2v} X(v_0/v_z) q\tau g D^2
\omega\sin(\chi)\cos(2\omega t+\chi).\nonumber\\
\end{eqnarray}
For $|v_0|\ll|v_z|$ the above equation is simplified as
\begin{eqnarray}
  \tilde{e}_{z\rm R,L}(t) &=& -\frac{4\pi v_0^2}{15vv_z^2} q\tau g D^2
\omega\sin(\chi)\cos(2\omega t+\chi).
\end{eqnarray}
The result also resembles that in the case $\alpha_{1,2}\ne0$ and $v_0=0$, with the only distinct feature in the doping dependence. The second harmonic response again vanishes in the $q\tau|D|\ll|\gamma|/v$ limit.

\subsection{Effects of Multiple Nodes in Weyl Semimetals}

So far, we have focused on the response from a Weyl node. However, in solids, there always exist multiple nodes~\cite{Neilsen1981}, $2n$ and $4n$ for centrosymmetric and time-reversal symmetric Weyl semimetals, respectively~\cite{Morimoto2016} ($n$ is a positive integer). A crucial difference is that by the time-reversal operation, one Weyl node is related to another with the same chirality. On the other hand, by the spatial inversion operation, the Weyl node is related to another with the opposite chirality. Therefore, it is expected that the photocurrent generally becomes nonzero for the Weyl semimetals with broken inversion symmetry, while the photocurrent cancels out between the nodes in presence of the spatial inversion symmetry.

As low energy models that take into account of the pair of nodes, we consider $4\times4$ Hamiltonian that consists of two nodes~\cite{Taguchi2016}. For the centrosymmetric Weyl Hamiltonian, we consider
\begin{eqnarray}
  H_P&=&v_0k_z\tau_z+\tau_z \{v(k_x\sigma_x+k_y\sigma_y)+v_zk_z\sigma_z\}\nonumber\\
  &&+g\tau_z (D_y\sigma_x-D_x\sigma_y)\nonumber\\
  &&+\frac{\alpha_1}2\sigma_z(k_x^2+k_y^2-2k_z^2)\nonumber\\
  &&+\frac{\alpha_2}2(k_zk_x\sigma_x+k_zk_y\sigma_y).
\end{eqnarray}
Here, $\sigma_\alpha$ and $\tau_\alpha$ ($\alpha=x,y,z$) are Pauli matrices for spin and chirality, respectively. For time-reversal symmetric Weyl Hamiltonian, we consider
\begin{eqnarray}
  H_T&=&v_0k_z\tau_z\pm\{v(k_x\sigma_x+k_y\sigma_y)+v_zk_z\sigma_z\}\nonumber\\
  &&+g\tau_z (D_y\sigma_x-D_x\sigma_y)\nonumber\\
  &&+\frac{\alpha_1}2\tau_z\sigma_z(k_x^2+k_y^2-2k_z^2)\nonumber\\
  &&+\frac{\alpha_2}2\tau_z(k_zk_x\sigma_x+k_zk_y\sigma_y).
\end{eqnarray}
In $H_P$ and $H_T$, we take the time-reversal operation operator as ${\cal T}=\sigma_y\tau_x K$ and spatial-inversion operator ${\cal P}=\tau_x$ in addition to $\bm k\to-\bm k$. The two models also have continuous rotation symmetry about the $z$ axis, if we assume the standard rotation operation for $\sigma_\alpha$ matrices, while no symmetry for the rotation about the $x$ or $y$ axis.

We note that the two Weyl nodes in $H_T$ have the same chirality. Hence, there exist at least two more nodes with the opposite chirality due to the Nielsen-Ninomiya theorem~\cite{Neilsen1981}. However, there is no symmetry relation between the two pairs of nodes. Therefore, in general, no complete cancellation occurs between any two pairs of different chiralities. Thus, here, we focus only on $H_T$ for the time-reversal symmetric case.

When the internodal scattering is negligible, the optical response in $H_P$ and $H_T$ is given as the sum of contributions from the two nodes in the Hamiltonian. Therefore, we see that the calculation based on $H_T$ and $H_P$ is consistent with the argument above.

For the second harmonic response, the sign of the response is independent of the chirality. However, $g$ has opposite sign between the pair nodes for both time-reversal and spatial inversion operations. Therefore, for both cases, in this mechanism, the second harmonic response vanishes otherwise the relaxation time becomes different between the pair nodes, or in a nonequilibrium setting such as chemical potential difference induced by chiral anomaly~\cite{Ebihara2016}.

\section{Discussion and Summary} \label{sec:summary}

In this paper, we have studied the effect of the emergent electric field in momentum space, $e_\mu(\bm k,t)$, in Weyl semimetals. We describe how the dynamical motion of the Weyl node induces the $e_\mu(\bm k,t)$ field, and that it resembles the magnetic field induced by the dynamical motion of the point charge in the Maxwell theory. It has been shown, in the previous paper~\cite{Ishizuka2016}, that the $e_\mu(\bm k,t)$ field induces a non-dissipative photocurrent in the adiabatic limit, i.e., $\tau\omega\gg1$, where $\omega$ is the frequency of incident light and $\tau$ is the relaxation time.

In this paper, we have studied how $e_\mu(\bm k,t)$ contributes to the electron transport in the regime of $\tau\omega\ll1$ by using the Boltzmann theory. We show that the non-dissipative photocurrent robustly remains in the Boltzmann theory. When the change in the fermion distribution function is mainly caused by the drift of electrons, i.e., when $|q\tau\bm D|\gg |\frac{g}v\bm D|$, the result is exactly the same with that in the adiabatic limit. The magnitude of the photocurrent, however, changes in the Boltzmann theory when the change in the fermion distribution function is dominated by the modification of the band structure, i.e., when $|q\tau\bm D|\ll |\frac{g}v\bm D|$. Nevertheless, even in this limit, the induced photocurrent is independent of $\tau$, namely, it is non-dissipative. This is a distinct feature from other mechanisms of the photocurrent in the Weyl semimetals~\cite{Taguchi2016,Chan2016_2}, as well as the photocurrent induced by the Berry curvature~\cite{Moore2010,Sodemann2015}. In these mechanisms, a change in the electron distribution function due to the drift of the electrons or by light irradiation plays key roles, and thus the photocurrent depends on the relaxation time. From the symmetry argument, we discuss that the photocurrent induced by our mechanism can generally be observed in the noncentrosymmetric Weyl semimetals such as TaAs~\cite{Huang2015,Lv2015,Sun2015,Xu2015,Weng2015}.

In addition to the non-dissipative current, we find that the second harmonic response (the a.c. current of $2\omega$ frequency) appears in the Boltzmann theory. This second harmonic response, however, cancels out between the Weyl nodes in the presence of either time-reversal or spatial-inversion symmetry. Therefore, to realize the second harmonic response, it is necessary to break both the time and inversion symmetries. A candidate experimental setup is to prepare a nonequilibrium state with a Fermi-level difference, $\mu_5$. The generation of $\mu_5$ using a circularly polarized light has recently been proposed in Ref.~\cite{Ebihara2016}. Our result suggests that when the light polarization is slightly distorted from the circular polarization, it gives rise to the second harmonic response of the electric current parallel to the separation axis of Weyl nodes.

\acknowledgements
The authors thank M. Ezawa and K. Kikutake for fruitful discussions. This work was supported by JSPS Grant-in-Aid for Scientific Research (No. JP24224009, No. JP26103006, No. JP26287088, No. JP15H05855, No. JP16H06717, and No. JP16J02240), from MEXT, Japan, and ImPACT Program of Council for Science, Technology and Innovation (Cabinet office, Government of Japan).

\appendix

\section{Boltzmann Theory for the Generalized Weyl Hamiltonian}\label{sec:boltzmann}

In this appendix, we elaborate on the formalism of the Boltzmann theory used in this paper. In Sec.~\ref{sec:boltzmann:formalism} we review a general formalism of the Boltzmann theory with time-dependent parameters. We elaborate on the general solution for the electron distribution function within the relaxation time approximation. In Sec.~\ref{sec:boltzmann:genWeyl}, we apply the theory in Sec.~\ref{sec:boltzmann:formalism} to the generalized Weyl Hamiltonian in Eq.~\eqref{eq:Hweyl}. We show a simple formalism to obtain the electron distribution function, which we use to calculate the optical responses in Sec.~\ref{sec:optresponse}.

\subsection{Boltzmann Equation} \label{sec:boltzmann:formalism}

In this section, we describe the Boltzmann theory for Hamiltonians with time dependent parameters $\boldsymbol \gamma(t)=(\gamma_1(t),\gamma_2(t),\cdots)$. Assuming a spatially uniform solution, the electron distribution at time $t$ is given by
\begin{eqnarray}
&&f(\bm k,\boldsymbol \gamma(t); t) =\nonumber\\
&&f(\bm k -\Delta t \dot{\bm k}(t-\Delta t),\boldsymbol \gamma(t-\Delta t); t-\Delta t) + \Delta t\dot f_\text{scatt},\nonumber\\
\end{eqnarray}
using the electron distribution function at $t-\Delta t$, $f(\bm k,\boldsymbol \gamma(t); t-\Delta t)$, where $\Delta t$ is a small time interval. Here, $\dot f_\text{scatt}$ is the scattering rate. Expanding up to the linear order in $\Delta t$, we get the Boltzmann equation
\begin{eqnarray}
  \partial_t f + \dot{\bm k}(t)\cdot\nabla_k f+\dot{\boldsymbol \gamma}(t)\cdot\nabla_\gamma f =  \dot f_\text{scatt},
\end{eqnarray}
where $\dot{\bm k}(t)=\frac{d}{dt}\bm k(t)$ and $\dot{\boldsymbol \gamma}(t)=\frac{d}{dt}\boldsymbol \gamma(t)$. Here, we abbreviate the arguments of the distribution function; $f=f(\bm k,\boldsymbol \gamma(t); t)$. We assume a small deviation from the equilibrium distribution function,
\begin{eqnarray}
f(\bm k,\boldsymbol \gamma(t); t) = f_0(\bm k,\boldsymbol \gamma(t) ) + g(\bm k,\boldsymbol \gamma(t); t),
\end{eqnarray}
where $f_0(\bm k,\boldsymbol \gamma(t); t)$ is the Fermi-Dirac distribution function, and $g(\bm k,\boldsymbol \gamma(t); t)\ll1$. Using the relaxation time approximation, $\dot f_\text{scatt}=-g/\tau$ with $\tau$ being the relaxation time, we get
\begin{eqnarray}
  \partial_t g + \dot{\vec k}(t)\cdot\nabla_{\vec k}f_0+\dot{\vec \gamma}(t)\cdot\nabla_\gamma f_0 =  -\frac{g}\tau.
\end{eqnarray}

Suppose both the incident electric field and the time-dependent parameters are monochromatic with the same frequency $\omega$, namely,
\begin{eqnarray}
  \bm D(t) &=& \bm D_+ e^{\rm i \omega t} + \bm D_- e^{-\rm i \omega t},\\
  \boldsymbol \gamma(t) &=& \boldsymbol\gamma_+ e^{\rm i \omega t} + \boldsymbol\gamma_- e^{-\rm i \omega t}.
\end{eqnarray}
Then, the solution of $g(\bm k,\boldsymbol\gamma(t);t)$ reads
\begin{eqnarray}
  g(\bm k,\boldsymbol\gamma(t);t)=-\tau f'_0(\varepsilon_{\vec k,\vec\gamma})\left[q\vec v\cdot \vec E(t)+\dot{\vec \gamma}\cdot\nabla_{\gamma}\varepsilon_{\vec k,\vec\gamma}\right].\nonumber\\
\end{eqnarray}
Here, for simplicity, we ignore the contribution from the Berry curvature $\bm b(\bm k)$, and only consider the leading order terms in $\tau$ by assuming $\omega\tau\ll1$. Using this result, the electron distribution reads
\begin{eqnarray}
  f(\bm k,\boldsymbol\gamma(t);t) &=& f_0(\bm k,\boldsymbol\gamma(t))+g(\bm k,\boldsymbol\gamma(t);t)\\
  &\sim& f_0(\bm k-q\tau\bm E,\boldsymbol\gamma(t)-\tau\cdot\dot{\boldsymbol\gamma}(t) ).\label{eq:boltzmann:formalism:relaxtime}
\end{eqnarray}
Therefore, the electron distribution can be approximated by the shift of parameters in the equilibrium distribution function.

\subsection{Generalized Weyl Hamiltonian}\label{sec:boltzmann:genWeyl}

For the generalized Weyl Hamiltonian in Eq.~\eqref{eq:Hweyl}, when $\alpha_{1,2}=0$, the Fermi-Dirac distribution function has the following relation 
\begin{eqnarray}
  f_0(\bm k,\boldsymbol\gamma(t) ) &=& f_0(\bm k-\frac1v\boldsymbol\gamma(t),\bm 0),\label{eq:boltzmann:genWeyl:dist}
\end{eqnarray}
where
\begin{eqnarray}
  \boldsymbol\gamma(t) =
  \left(
  \begin{array}{c}
     \gamma_x \\
     \gamma_y \\
      0
  \end{array}
  \right)=
  \left(
  \begin{array}{c}
    -gD_y \\
     gD_x \\
      0
  \end{array}
  \right).\label{eq:boltzmann:genWeyl:shift}
\end{eqnarray}
This relation does not hold for $\alpha_{1,2}\ne0$. However, in this section, we discuss that the distribution function can be approximated by Eq.~\eqref{eq:boltzmann:genWeyl:dist} when $\alpha_{1,2}k_F\ll v, v_z$ and $|\boldsymbol \gamma| \ll vk_F$.

From the above observation, we assume that the electron distribution function can be approximated as
\begin{eqnarray}
  f_0(\bm k,\boldsymbol\gamma) \sim f_0(\bm k-\bm{\delta k},\bm0), \label{eq:boltzmann:genWeyl:assum}
\end{eqnarray}
where
\begin{eqnarray}
  \bm{\delta k} =
  \left(
  \begin{array}{c}
    \delta k_x \\
    \delta k_y \\
      0
  \end{array}
  \right).
\end{eqnarray}
Here $\delta k_x=\delta k_x(\gamma_x)$ and $\delta k_y=\delta k_y(\gamma_y)$ are a real function of $\gamma_x$ and that of $\gamma_y$, respectively. We derive $\bm{\delta k}$ from the relation on the energy eigenvalue $\varepsilon(\bm k,\boldsymbol\gamma)$,
\begin{eqnarray}
   \varepsilon(\bm k,\boldsymbol\gamma)=\varepsilon(\bm k-\bm{\delta k},\bm 0).\label{eq:boltzmann:genWeyl:eRel}
\end{eqnarray}

Expanding Eq.~\eqref{eq:boltzmann:genWeyl:eRel} to the first order in $\boldsymbol \gamma$ and $\bm{\delta k}$, we obtain relations
\begin{subequations}
  \begin{align}
  v_x(\bm k,\bm0)\delta k_x&=R_x(\bm k,\bm0)\gamma_x,\\
  v_y(\bm k,\bm0)\delta k_y&=R_y(\bm k,\bm0)\gamma_y.
  \end{align}\label{eq:boltzmann:genWeyl:selfcons}%
\end{subequations}
Here, $R_x=R_x(\bm k,\boldsymbol\gamma)$ and $R_y=R_y(\bm k,\boldsymbol\gamma)$ are the parameters of the generalized Weyl Hamiltonian in Eq.~\eqref{eq:Rweyl}, and $v_\alpha(\bm k,\boldsymbol\gamma)=\partial_\alpha\varepsilon(\bm k,\boldsymbol\gamma)$ ($\alpha=x,y,z$) are the group velocities. Solving Eq.~\eqref{eq:boltzmann:genWeyl:selfcons}, we find
\begin{eqnarray}
  \delta k_x&=& \left(\frac{g}v+{\cal O}(\frac{g}v\frac{\alpha_{1,2}}v k_\alpha)\right)D_y,\\
  \delta k_y&=&-\left(\frac{g}v+{\cal O}(\frac{g}v\frac{\alpha_{1,2}}v k_\alpha)\right)D_x.
\end{eqnarray}
Therefore, to the lowest order, the distribution function of the generalized Weyl Hamiltonian is approximated by Eq.~\eqref{eq:boltzmann:genWeyl:assum}, with
\begin{eqnarray}
  \delta k_x&=& \frac{g}v D_y,\\
  \delta k_y&=&-\frac{g}v D_x.
\end{eqnarray}
This distribution function gives the same form with Eq.~\eqref{eq:boltzmann:genWeyl:shift}. Therefore, we approximate the electron distribution function as in Eq.~\eqref{eq:boltzmann:genWeyl:dist}.

Combining this result with the result of the relaxation time approximation in Eq.~\eqref{eq:boltzmann:formalism:relaxtime}, we get
\begin{eqnarray}
  f(\bm k,\boldsymbol\gamma(t);t) &\sim& f_0(\bm k-q\tau\bm D-\frac1v (\boldsymbol\gamma-\tau\cdot\dot{\boldsymbol\gamma}),\bm0)\\
  &\sim& f_0(\bm k-q\tau\bm D-\bm{\delta k},\bm0).\label{eq:boltzmann:genWeyl:edist}
\end{eqnarray}
In the second line, we ignore the $\dot{\boldsymbol\gamma}$ term since we assume $\tau\omega\ll1$. In the first argument of $f_0$ in the second line, the second term comes from the drift of electrons by the external field, and the third term from the deformation of the electron orbitals by the external electric field.

In the last, we estimate the order of the magnitude of the second and third terms in the first argument of $f_0$ in Eq.~\eqref{eq:boltzmann:genWeyl:edist}. Assuming $v=10^{-29}$ Jm, $g=10^{-29}$ Jm/V, $\omega=10^{13}$, and $\tau=10^{-15}-10^{-14}$s, we get
\begin{eqnarray}
  |\frac{g}v\bm D|&\sim& 10^0D,\\
  q\tau|\bm D|&\sim& 10^0D-10^1D.
\end{eqnarray}
Therefore, in most cases, the contribution from $q\tau\bm D$ would be dominant. It is, however, possible that the $|g\bm D/v|$ term gives an observable contribution to the change in the electron distribution function. Therefore, in the main text, we consider two limits: $|g\bm D/v|\gg q\tau|\bm D|$ and $|g\bm D/v|\ll q\tau|\bm D|$ for completeness.

\end{document}